\documentclass[mathleft]{an}
\usepackage{graphicx}
\usepackage{times}
\begin{document}

\Pagespan{789}{}
\Yearpublication{2006}%
\Yearsubmission{2005}%
\Month{11}%
\Volume{999}%
\Issue{88}%

\title{Extragalactic X--ray Surveys: AGN physics and evolution} 
        
\author{Andrea Comastri\inst{1}\fnmsep\thanks{Corresponding author
  \email{andrea.comastri@oabo.inaf.it}\newline}
\and  Marcella Brusa\inst{2}
}
\titlerunning{Extragalactic X--ray Surveys}
\authorrunning{A. Comastri \& M. Brusa}
\institute{
INAF-Osservatorio Astronomico di Bologna, Via Ranzani 1, I-40127 Bologna, Italy
\and 
MPE, Giessenbachstrasse 1, D-85748 Garching, Germany}

\received{3 Sep 2007}
\accepted{          }
\publonline{later}


\abstract{%
We review the most important findings on AGN physics and cosmological evolution 
as obtained by  extragalactic X--ray surveys and associated 
multiwavelength observations. We briefly discuss the perspectives for future enterprises and
in particular the scientific case for an extremely deep (2--3 Ms) XMM survey} 

\maketitle

\section{Introduction}

Since their launch in 1999, both XMM--{\it Newton} and {\it Chandra} have performed a large 
($>$ 30) number of surveys covering a wide fraction of the area vs. depth plane 
(see fig~1 in Brandt \& Hasinger 2005). 
Thanks to vigorous programs of multiwavelength follow-up observations 
which, since a few years, have became customary, our understanding of AGN evolution 
has received a major boost.
The high level of completeness in redshift determination for a large number 
of X--ray selected AGN (up to a few thousands)
has made possible 
a robust determination of the luminosity function  and evolution of unobscured 
and mildly obscured AGN 
(Ueda et al. 2003; Hasinger et al. 2005; La Franca et al. 2005), which
turned out to be luminosity dependent:
the space density of bright QSO ($L_X > 10^{44}$ erg s$^{-1}$) peaks at z$\sim$ 2--3, 
to be compared with the z$\sim$0.7--1 peak of lower luminosity  Seyfert galaxies.

The fraction of obscured AGN is also strongly dependent from the 
X--ray luminosity (Ueda et al. 2003; La Franca et al. 2005). 
Even though the shape and the normalization 
of the function describing the obscured fraction vs. luminosity 
is still matter 
of debate, it is clear that absorption is much more common at low X--ray luminosities. 
Such a trend has been observed 
also in the optical (Simpson 2005) and in the near infrared (Maiolino et al. 2007) 
and may  be linked to the AGN radiative power which is able to ionize and expel
gas and dust from the nuclear regions.
More debated is the claim of an increase of the obscured fraction 
towards high redshifts. First suggested by La Franca et al. (2005), it has been 
confirmed by Treister \& Urry (2006), while it is not required 
in the models discussed by Ueda et al. (2003) and Gilli et al. (2007). 

A redshift dependence of the obscuring fraction would be naturally explained 
in the current framework of AGN formation and evolution (see, for example, 
Hopkins et al. 2006):
the anti--hierarchical behaviour observed in AGN e\-vo\-lution (similar to that observed in 
normal galaxies), along with several other evidences, suggests that super\-massive 
bl\-ack holes (SMBH) 
and their host galaxies co--evolve and that their formation and evolution are most likely 
different aspects of the same astrophysical problem. 
At early times large quantities of cold gas were available 
to efficiently feed and obscure the growing black holes. Later on, the ionizing nuclear flux
is able to "clean" its environment appearing as an unobscured QSO. 
However, the picture sket\-ched above is likely to be much more complicated, depending 
on many other parameters (such as the BH mass, the Eddington ratio, the QSO duty cicle) 
and may not necessarily result in an increasing fraction of obscuration towards 
high redshifts.

It should also be remarked that sensitive X--ray observations 
are highly efficient to unveil weak and/or elusive accreting black holes 
which would be missed, or not classified as such, by surveys at longer wavelengths.
Among them XBONG (Comastri et al. 2002), sources with high X--ray to optical 
flux ratio (Fiore et al. 2003), Extremely Red Objects (Brusa et al. 2005), 
Sub Millimeter Galaxies (Alexander et al. 2005). 
Although they are probably not representative of the sources of the 
X--ray background their study has allowed us to better understand the
physics of accreting black holes.

Finally, the coverage of several {\it Chandra} and XMM fie\-lds has
allowed to uncover several redshifts spikes in the distribution of X--ray 
sources (Gilli et al. 2005) and  demonstrated that AGN may be used 
as reliable tracers of the large scales structures. The underlying 
large scale structure may play an important role in triggering AGN activity.
According to the analysis of a sample of X--ray selected AGN in the
AEGIS survey, Georgakakis et al. (2006) suggest that $z \sim$ 1 AGN
are more frequent in dense environments.

\section{Open Questions} 

The most important achievements, briefly outlined above, were obtained combining 
data from both deep and large area surveys. The former were conceived to reach extremely 
faint fluxes at the expenses of a reduced sky coverage, while for the latter  
the search of a trade-off between area and exposure time is the main driver.
Both the approaches have their own scientific goals which cannot be directly 
compared, but rather considered as a necessary sinergy.
The de\-epest surveys in the {\it Chandra} Deep Fields North and South (CDF) have reached 
extremely faint fluxes over a relatively small ($\sim$ 400 arcmin$^2$ each) portion of the 
sky and represent an unique resource for the study of the 
faint end of the luminosity function and the discovery of distant and obscured AGN.
The search for most luminous quasars and the study of the clustering of 
the AGN population is uniquely pursued by large area surveys.
At present the most relevant projects in terms of area, X--ray flux and
multiwavelength follow--up programs are the XMM--COSMOS 2 sq. degree survey 
(Hasinger et al. 2007, Cappelluti et al. 2007), the {\it Chandra}--AEGIS survey 
(Nandra et al. 2005) and the Extended CDFS (Lehmer et al. 2005).  

New results always bring new questions. 
An exhaustive review of open problems is beyond the purposes of the present paper and we
highlight what we consider the most pressing questions and which of them could
be answered by future XMM--{\it Newton} surveys.

As far as SMBH are most directly concerned, attention is now focused on a 
few key questions:
\par
$\bullet$ How is star formation and gas accretion linked ? 
How strong is the relative feedback ? What is the role of the environment in 
triggering AGN activity ? 
\par
$\bullet$ How many highly obscured AGN are still missing, how much do they contribute  
to the accretion history and SMBH mass budget in the Universe?  
\par
$\bullet$ How strong is the dependence of obscuration upon luminosity and redshift ? Are emission
K$\alpha$ lines common also at high redshift ?
\par
$\bullet$ What is the space density of high redshift ($>$ 3) qua\-sars ? How many of them
are obscured ?    

Some of these topics are being addressed with the available 
multiwavelength data.

\subsection{``Missing'' AGN} 
The X--ray background (XRB) below 5--6 keV (Worsley et al. 2005),  
has been almost completely resolved into single 
sources, the large majority of them being obscured by substantial 
amounts of gas (up to 10$^{24}$ cm$^{-2}$; i.e. in the Compton thin regime), 
thus nicely confirming the predictions of AGN synthesis models. 
The {\it unresolved} XRB fraction, which provides an integral constraint
to the number of ``missing" A\-GN, increases with energy and is close
to 100\% at the 30 keV peak.
A good fit is obtained if a population of heavily obscured 
Compton Thick (CT; $N_H > 10^{24}$ cm$^{-2}$) AGN whose  space density 
is of the same order of that of Compton thin is added. 
Such an estimate is model dependent and probably subject to large uncertainties 
however, while  abundant in the local Universe, only a handful of CT AGN are known
at cosmological distances (see Comastri 2004 for a review). 
Either CT AGN are numerous also at high z and the bulk of population is 
still undetected, or they represent a "local" phenomenon 
possibly associated with a "final" phase in the evolution of nuclear 
obscuration. In the latter hypothesis the 30 keV peak of the XRB 
would originate in different, yet unknown, sources.
According to synthesis models the fraction of CT AGN 
is expected to steeply increase just beyond the present limits (Fig.~1) 
and indeed the deepest search for CT AGN in the X--ray band 
(Tozzi et al. 2006) uncovered a number of candidates which is fairly consistent 
with the model predictions, supporting the hypothesis that CT AGN are too 
faint to be directly detected in the X--ray band.  

\begin{figure}[!t]
\includegraphics[width=80mm,height=80mm]{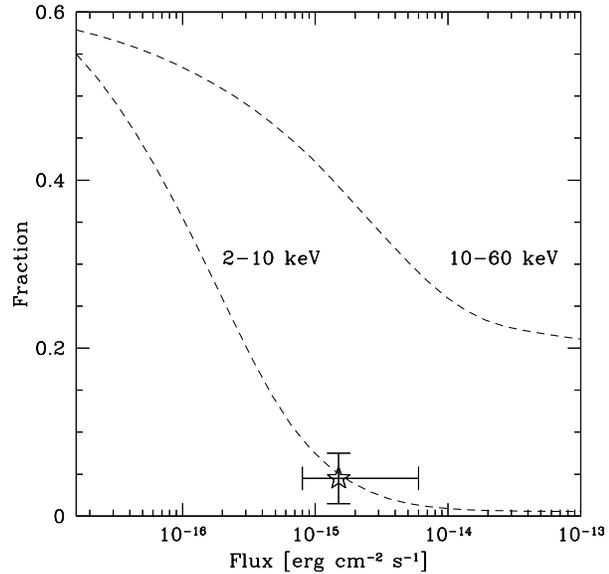}
\caption{The predicted fraction of CT AGN 
(Gilli et al. 2007) as a function of the X--ray flux in the 2--10 keV 
band along with the estimate of Tozzi et al. (2006). The model predicted 
fraction in the 10--60 keV band is also reported.}
\label{label1}
\end{figure}

Alternative techniques, based on the selection via infrared and radio data, seem
promising (i.e. Martinez Sansigre et al. 2005). Indeed Fiore et al. (2007) and 
Daddi et al. (2007) were able to recover the "missing" obscured CT AGN population
making use of 24$\mu$m {\it Spitzer} observations and
deep near--infrared and optical data.
Stacking of the X--ray counts of CT candidates selected on the basis of an infrared 
excess and individually undetected in the CDF Ms exposures, revealed a strong signal in the 
hard ($\sim$ 4--8 keV) band which imply, at their average redshift ($z\sim$2), 
absorption column densities in the CT regime.
The census of CT and obscured AGN population bears important con\-se\-quen\-ces 
for the study of the assembly and evolution of SMBH. 
For example the mass function of Compton thin AGN, estimated from the luminosity function of
X--ray selected AGN, falls short by a factor 2 from that of "relic" SMBH 
in local bulges (Marconi et al. 2004). The candidate CT AGN population 
selected in the infrared would have the right size to reconcile the 
two SMBH mass function estimates.
It has also been suggested that the absorption by CT matter of
high energy photons may be able to efficiently heat the surrounding 
material through Compton scattering (Daddi et al. 2007). 
CT sources may thus play a key role in the
AGN feedback eventually leading to the quenching of star formation. 

\subsection{High-redshift Quasars} 

The space density and evolution of high redshift QSO is still an open issue. 
In the optical band, mainly thanks to the SDSS, their luminosity function
is relatively well known up to $z \sim$ 6. Although the highest redshift 
QSO are known to be X--ray emitters (Vignali et al. 2003) the lack of sensitive and
wide enough surveys with a fairly homogenous and complete coverage has not allowed
to obtain reliable estimates of high redshift X--ray selected QSO.
The XMM-COSMOS survey is starting to fill this gap.
So far seven spectroscopically confirmed QSO at $z >$ 3 were found abo\-ve a flux
of $2\times 10^{-15}$ cgs (where the sky coverage is flat over the entire
2 sq degs area).
By fully exploiting the multicolor optical and near infrared
diagrams several candidate $z >$ 3 QSO
are found at the same limiting flux, bringing the total number 
of candidate in the COSMOS field to $\sim$ 40 (Brusa et al.in preparation).
The upper and lower limits of the $z >$ 3  QSO counts 
are reported in Fig.~2, along with the number counts of
the overall X-ray source population from a compilation of
different surveys (Cappelluti et al. 2007).
The two magenta lines show the contribution of $z >$3 QSO
predicted by XRB synthesis models (Gilli, Comastri \& Hasinger 2007) 
upon two different assumptions on their high redshift evolution:
an exponential decline at $z >$ 3 as parameterized by Schmidt et al. (1995)
or a constant space density.
Although the actual number of $z >$ 3 QSO in the COSMOS field suffers
from large uncertainities, the present data would suggest that
a constant space density at high redshift is ruled out.

\begin{figure}[!t] 
\includegraphics[width=80mm,height=80mm]{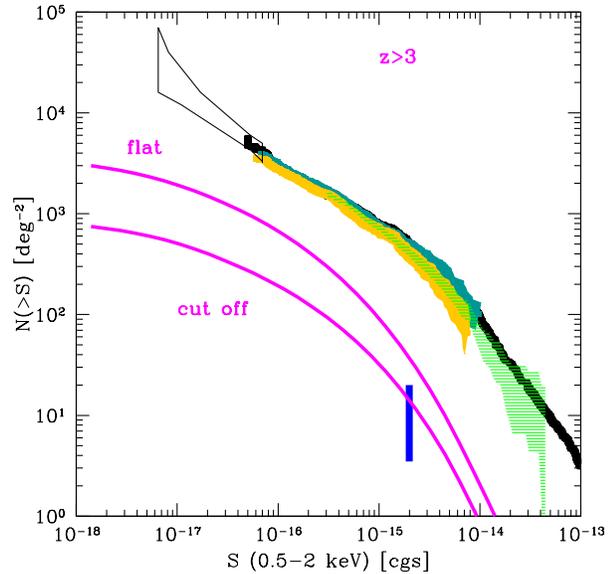}
\caption{The observed space density of $z>$ 3 quasars 
in the COSMOS field (blue vertical line) compared with the predicted number counts  
(Gilli et al. 2007) under two different hypotheses (see text)}
\label{label1}
\end{figure}


\section{Perspectives} 

In order to properly address the key questions a better sensitivity, over a large 
sky area as well as an extended energy range than obtained to date, are required.
While in principle one would need all of them at once, in  practice 
this is impossible.
An ultradeep (2--3 Ms) hard X--ray survey with XMM--{\it Newton} would allow 
the scientific community to address with an unprecedent accuracy 
several of the "key" questions outlined above and in particular those concerning
the most obscured "missing" AGN and the X--ray spectral properties 
(absorption, emission lines, etc) of distant sources.
The large collecting area and hard X--ray sensitivity of the 
imaging detectors onboard XMM have not yet been pushed to their limits
(the deepest XMM survey in the Lockman Hole reached an exposure time of $\sim$ 700 ks). 
According to our calculations (see also Carrera et al. this volume) 
the {\tt EPIC} camera would not be confusion limited 
down to fluxes of the order of $3-5 \times 10^{-16}$ cgs 
in the 5--10 keV band corresponding to a factor 3--4 deeper than the 
present limits (Fig.~3).   

\begin{figure*}
\includegraphics[width=160mm,height=120mm]{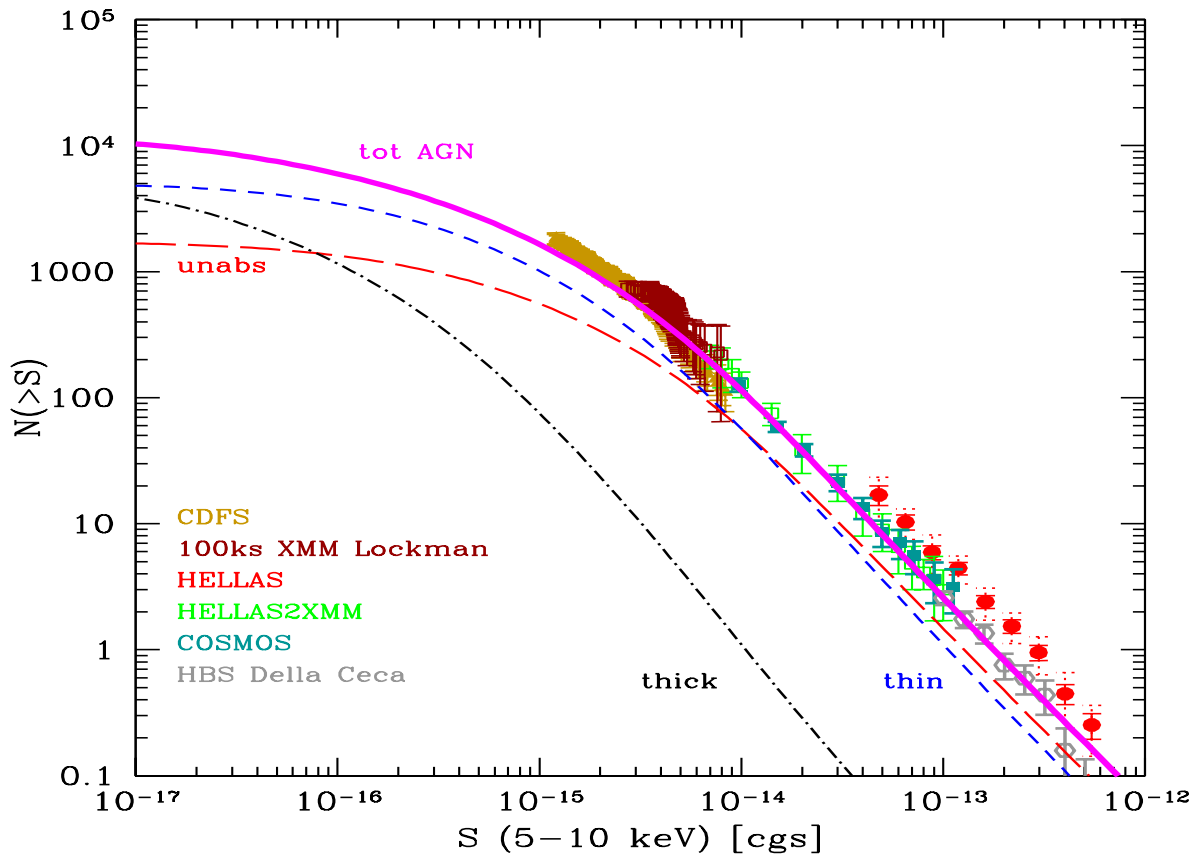}
\caption{The 5--10 keV counts, from a compilation of several 
surveys as labeled, compared with the Gilli et al 2007 model predicitons
for unobscured (red line), Compton thin (blue line), Compton Thick 
(black line), and total AGN (magenta line)} 
\label{label1}
\end{figure*}

The model predicted logN--logS of CT AGN, in the 5--10 keV energy range, has
a steep slope down to faint fluxes and thus a deeper exposure would allow 
to detect several new candidates, presumably at moderate to high redshifts, responsible
for the unresolved XRB. Given that the Gil\-li et al. (2007) model 
reproduce the XRB level measured by HEAO1 (fig.~4), the predicted counts 
should be considered as a lower limit.
The very shape and normalization of the XRB spectrum in the 
$\approx$ 5--15 keV band is, at present, quite uncertain.
On the one hand, recent {\tt BeppoSAX} (Frontera et al. 2007); {\tt INTEGRAL} 
(Churazov et al. 2007) and {\tt Swift} (Ajello et al. in preparation) observations 
have unambiguosly demonstrated that the XRB intensity above 15 keV 
is close to that measured by HEAO1 (fig.~4).
On the other hand, according to a detailed analysis of deep {\it Chandra} field 
(Hickox \& Markevitch 2006) the XRB level, below about 5 keV,
is some 30\% higher than the extrapolation of higher energy data. 
Given that the summed contribution of faint {\it Chandra} sources 
in the 1--8 keV band is already exceeding, at the face value, 
the HEAO1 level, it may well be possible that 
a much larger fraction than that quoted by Worsley et al. (2005) has been already resolved.
Alternatively, a sizable population of extremely hard sources, 
appearing only above 5--6 keV, and not accounted for in the XRB synthesis 
model, are providing a significant contribution.

\begin{figure*}
\includegraphics[width=160mm,height=125mm]{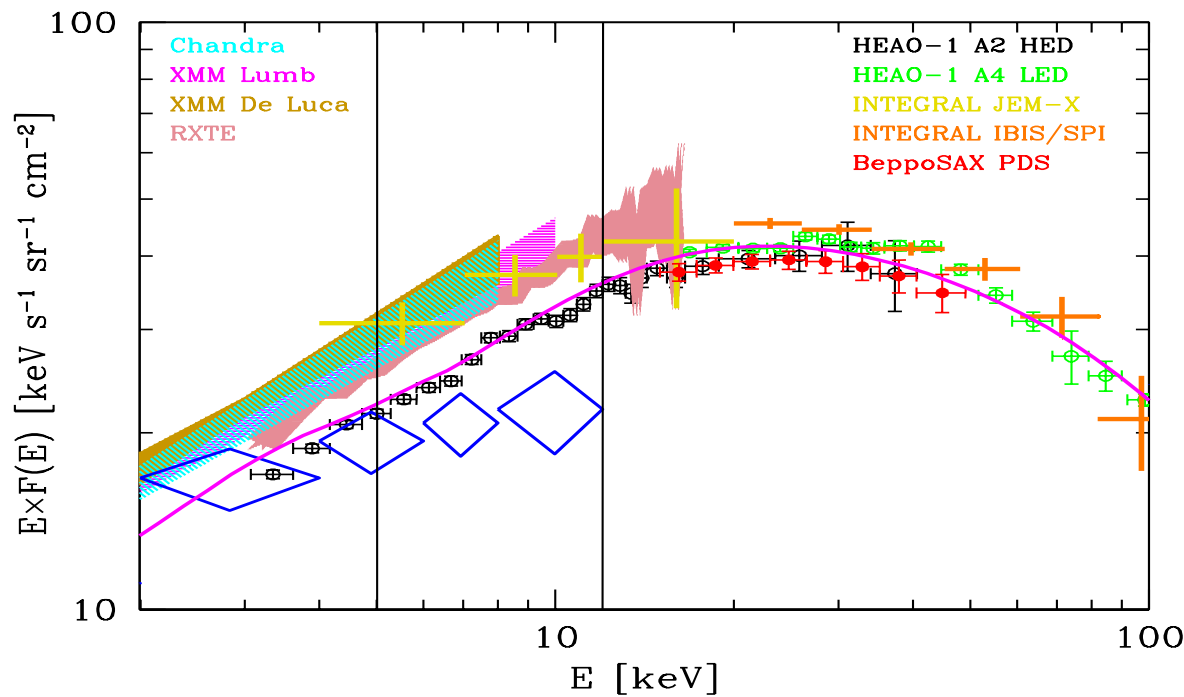}
\caption{A selection of XRB measurements, including the resolved fraction 
in deep fields (blue squares) and the best fit (magenta line) AGN synthesis model of 
Gilli et al. (2007). The vertical lines mark the 5--12 keV energy range}
\label{label1}
\end{figure*}

\par
Interesting enough 
fully, 4$\pi$, covered AGN with a  negligible fraction of 
scattered flux in the soft X--rays 
may provide a relevant contribution only above 5 keV or even more depending
from their redshift and column density. 
 A few examples of these sources in the 
local Universe were recently discovered with Suzaku 
(Ueda et al. 2007; Comastri et al. 2007). 
Should a population of heavily obscured, 
fully covered AGN be abundant at moderate to high redshifts 
they might be discovered only by sensitive surveys above 5 keV. 
\par
A deep XMM survey 
would allow to collect good quality X--ray spectra for a large number 
of sources. The obvious advantages of an improved spectral analysis concern 
a better determination of the absorption column densities especially 
at high redshifts where the determinations based on the Hardness Ratios 
suffer of the largest uncertainties. 
\par
The studies based on the stacking of X--ray detected sources would also benefit from 
an improved counting sta\--ti\--stic. 
Such a technique has been widely adopted for different purposes. Among them,
the study of the intensity and profile of iron K line in the average 
spectrum of faint AGN. Convincing evidence for the 
presence of iron emission up to high ($z \sim$ 2--3) redshifts 
has been presented by Brusa et al. (2005), while Streblyanska et al. (2005)
reported the discovery of a broad line with an extended red wing in the 
average spectra of both Type 1 and Type 2 AGN. 
The line profile brings unique information on the SMBH spin and is the 
signature that General Relativistic effects are at work in the innermost 
regions of AGN. 
\par
An ultradeep X--ray survey is also well suited to search for 
high redshift AGN. As shown in Fig.~2, the differences in the counts
for a model with and without a cut--off in the evolution at high redshift 
increase towards faint fluxes. 
Moreover, the search for accreting SMBH among various classes of objects selected on the 
basis of longer wavelength properties (see $\S$ 2.1) can be 
exploited in more detail. It is worth mentioning that the selection 
on the basis of an extreme infrared to optical color
seems quite promising to pick up the first ($z >$ 6) QSO 
(Koekemoer et al. 2004) 
\par
The merit of an ultradeep XMM survey should be jud\-ged 
in the context of multiwavelength deep surveys and its scientific 
return would be maximized if it is carried out in the best studied fields.
The excellent coverage of the Chandra Deep Field South 
with both present and future (i.e. AL\-MA and SKA) facilities, 
along with the already planned additional 1 Ms of DDT {\it Chandra} data, 
are making the CDFS a legacy field for the years to come.
The so--called co--evo\-lu\-tion of AGN and host galaxies and their
mutual feedback 
represents one of the most important achievements in the field of
observational cosmology. It is by now clear that further advances 
can be obtained only by joining the efforts of people working in 
apparently different fields and at different wavelengths 
(crudely: optical near--IR for the galaxy evolution, X--ray surveys for AGN).  
Deep XMM observations will, at the same time, benefit and enhance 
the return of the many common  scientific goals.

%

\acknowledgements
We thank Roberto Gilli for useful discussions and for help with Fig. 2. 
We acknowledge financial contribution from contract ASI--INAF I/023/05/0,
PRIN--MUR grant 2006--02--5203, and the BMWI/DLR FKZ 50OX 0002 grant.


\begin{thebibliography}{}
  \bibitem{} Alexander D.M. et al.: 2005, Nature 434, 738
  \bibitem{} Brandt W.N. \& Hasinger.: 2005, ARA\&A 43, 829
  \bibitem{} Brusa M., Comastri A., Daddi E. et al.: 2005, A\&A 
  \bibitem{} Brusa M., Gilli R., Comastri A.,: 2005, ApJ 621, L5
  \bibitem{} Cappelluti N., et al.: 2007, ApJS 172, 341
  \bibitem{} Comastri A., et al.: 2002, ApJ 571, 771 
  \bibitem{} Comastri A.: 2004, Ap\&SS Library 308, 245, Dordrecht Kluwer
  \bibitem{} Comastri A., et al.: 2007, astro-ph/0704.1253 
  \bibitem{} Churazov E., et al.: 2007, A\&A 467, 529 
  \bibitem{} Daddi E., et al.: 2007, ApJ in press astro-ph/0704.2832
  \bibitem{} Fiore F., Brusa M., Cocchia F. et al.: 2003, A\&A 409, 79
  \bibitem{} Fiore F., et al.: 2007, ApJ in press astro-ph/0704.2864
  \bibitem{} Frontera F., et al.: 2007, ApJ 666, 86
  \bibitem{} Georgakakis A., et al.: 2007, ApJ in press (astro--/ph/0607274)
  \bibitem{} Gilli R., et al.: 2005, A\&A 430, 811
  \bibitem{} Gilli R., Comastri A., Hasinger G.: 2007, A\&A 463, 79
  \bibitem{} Hasinger G. et al.: 2007, ApJS 172, 295 
  \bibitem{} Hickox R.C. \& Markevitch M.: 2006, ApJ 645, 95 
  \bibitem{} Hopkins P.F. et al.: 2006, ApJS 163, 1
  \bibitem{} Koekemoer A., et al.: 2004, ApJL 600, 123 
  \bibitem{} La Franca F., Fiore F., Comastri A., et al.: 2005, ApJ 635, 86
  \bibitem{} Lehmer B.D., et al.: 2005, ApJS 161, 21 
  \bibitem{} Maiolino R., et al.: 2007, A\&A 468, 979
  \bibitem{} Marconi A., et al.: 2004, MNRAS 351, 169
  \bibitem{} Martinez-Sansigre A., et al.: 2005, Nature 436, 666
  \bibitem{} Nandra K., et al.: 2005, MNRAS 356, 568
  \bibitem{} Schmidt K., Schneider P., Gunn J.E.: 1995, AJ 110, 68
  \bibitem{} Simposon C.: 2005, MNRAS 360, 565 
  \bibitem{} Streblyanska A., et al. : 2005, A\&A 432, 395 
  \bibitem{} Tozzi P., et al.: 2006, A\&A 451, 457
  \bibitem{} Treister E. \& Urry C.M.: 2006, ApJ 652, L79
  \bibitem{} Ueda Y., et al.: 2003, ApJ  598, 886
  \bibitem{} Ueda Y., et al.: 2007, ApJ  664, 79
  \bibitem{} Vignali C., Brandt W.N., Schneider D.P.: 2003, AJ 125, 433
  \bibitem{} Worsley M.A., et al.: 2005, MNRAS 357, 1281
\end{thebibliography}
\end{document}